\documentclass[aps,prl,twocolumn,superscriptaddress,showpacs,amsmath,floatfix,citeautoscript]{revtex4-2}

\usepackage[utf8]{inputenc}
\usepackage{graphicx}
\usepackage{mathtools}
\usepackage{xcolor}
\usepackage{amsmath}
\usepackage{amssymb}
\usepackage{xfrac}
\usepackage{soul}
\usepackage{mhchem}

\begin{document}
	
	\title{Insulator-to-Metal Transitions Driven by Quantized Formal Polarization Mismatch}

	\author{Hongsheng Pang}
	\affiliation{Laboratory of Quantum Information, University of Science and Technology of China, Hefei 230026, China}
	
	\author{Lixin He}
	\email{helx@ustc.edu.cn}
	\affiliation{Laboratory of Quantum Information, University of Science and Technology of China, Hefei 230026, China}
	\affiliation{Institute of Artificial Intelligence, Hefei Comprehensive National Science Center, Hefei, 230088, China}
	\affiliation{Hefei National Laboratory, University of Science and Technology of China, Hefei 230088, China}
	
	\date{\today}

	\begin{abstract}
		
We propose a mechanism for insulator-to-metal (IM) transitions driven by the mismatch of quantized formal polarization (QFP), a symmetry-protected bulk invariant. For a material with a low-symmetry insulating phase and a high-symmetry phase that allow distinct QFPs, any continuous path connecting them while preserving the symmetry of the low-symmetry phase must inevitably pass through an IM transition. The reason is that QFP remains invariant along any gapped symmetry-preserving evolution, whereas the high-symmetry phase requires a different QFP, which can only be accommodated by gap closing. First-principles calculations on two representative systems, two-dimensional $\ce{InPS3}$ and three-dimensional $\ce{CdBiO3}$, confirm this mechanism. Our results establish QFP mismatch as a general symmetry constraint on phase evolution and reveal a new route to symmetry-driven IM transitions in high-symmetry materials.
		
	\end{abstract}
	
	\maketitle
	
	
The insulator-to-metal (IM) transition is one of the most fundamental phenomena in condensed matter physics, with profound implications for material properties and device applications~\cite{RevModPhys.70.1039}. A wide variety of mechanisms can drive such transitions, including electronic correlations and external perturbations such as pressure or electric fields. In topological systems, changes in topological invariants can trigger topological quantum phase transitions, which are often accompanied by band-gap closing and may consequently lead to IM transitions~\cite{RevModPhys.82.3045}.

In crystals, electric polarization is constrained by the generalized Neumann principle~\cite{pang2025}. Within the modern theory of polarization~\cite{vanderbilt1993electric,vanderbilt2018berry}, this leads to quantized polarization in high-symmetry solids, historically referred to as {\it formal} polarization.
For a long time, quantized formal polarization (QFP) was regarded as a nominal quantity, and its physical significance was largely overlooked. However, recent studies have revealed that QFP has direct physical consequences, including fractional quantum ferroelectricity (FQFE)~\cite{Ji2024,Yu2025,pang2025quantized}. Moreover, interfaces between materials with distinct QFPs can host remarkable interfacial phenomena~\cite{hao2025interface,PhysRevB.80.241103,pang2025quantized}.
These developments suggest that QFP is not merely a formal label, but a physically meaningful bulk invariant with profound consequences for material behavior.
	
In this Letter, we propose a mechanism for IM transitions driven by QFP, a symmetry-protected bulk invariant. For a material with a low-symmetry phase L and a high-symmetry phase H that allow distinct QFPs, any continuous path connecting them while preserving the symmetry of L must inevitably undergo an IM transition, provided that L is insulating. This follows from the fact that QFP remains invariant along any gapped, symmetry-preserving evolution, whereas the higher symmetry of H requires a different allowed value. The resulting mismatch cannot be resolved along a fully insulating path and can only be accommodated through gap closing. Thus, the emergence of a metallic state is a direct consequence of QFP mismatch.
We demonstrate this mechanism in both two and three dimensions using first-principles calculations on two representative systems, namely the two-dimensional material $\ce{InPS3}$ and the three-dimensional compound $\ce{CdBiO3}$. Our work establishes QFP mismatch as a general symmetry constraint on phase evolution and reveals a new route to symmetry-driven IM transitions in high-symmetry materials.

	
First-principles calculations are performed using the ABACUS package~\cite{li2016,chen2010} with the Heyd--Scuseria--Ernzerhof (HSE) functional~\cite{hse}. Optimized norm-conserving Vanderbilt (ONCV) pseudopotentials~\cite{Hamann2013ONCV} from the SG15 set~\cite{Scherpelz2016SG15} are employed, with spin-orbit coupling fully included. Numerical atomic orbital (NAO) basis sets~\cite{Linpz2023} are used throughout. Electric polarization, band gaps, fat-band structures, and projected density of states (PDOS) are calculated using the \textsc{PYATB} package~\cite{jin2023}, based on tight-binding Hamiltonians constructed from self-consistent ABACUS calculations. Standard symmetry-operation matrices and Wyckoff positions are obtained from the Bilbao Crystallographic Server~\cite{bilbao}. Additional computational details are provided in the Supplemental Material (SM)~\cite{SM}.

%
	
\begin{figure}[tbp]
		\centering
		\includegraphics[width=0.45\textwidth]{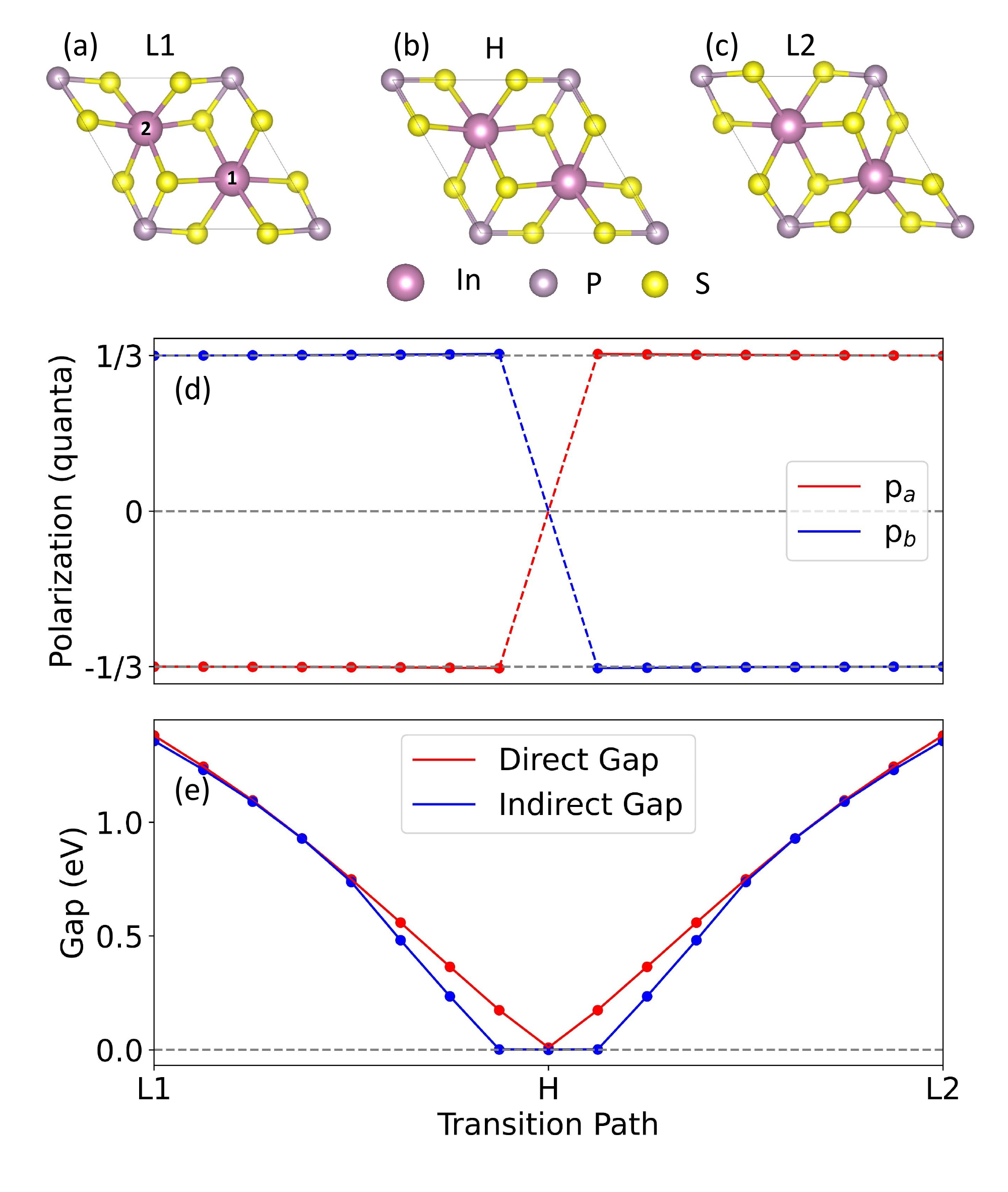}
		\caption{(a)--(c) Top views of $\ce{InPS3}$ in different phases: (a) L1 phase, (b) H phase, and (c) L2 phase. The two In atoms are labeled for distinction.  (d) Variation of polarization along the transition pathway, with the dashed line denoting the metallic state. (e) Variation of direct (red line) and indirect band gap (blue line) along the transition pathway.}
		\label{fig:inps3_neb}
	\end{figure}

The crystal structures of the two-dimensional material $\ce{InPS3}$ are shown in Fig.~\ref{fig:inps3_neb}(a)-(c). The leftmost structure, denoted as L1, crystallizes in space group $P312$ and belongs to the point group $D_3$. The corresponding Wyckoff positions are listed in Table S2 of the SM~\cite{SM}. The P atoms occupy the $2g$ positions in the plane. 	
The two In atoms occupy the symmetry-related fractional positions $1d\,(1/3,\,2/3,\,1/2)$ and $1f\,(2/3,\,1/3,\,1/2)$, respectively, while the S atoms occupy the high-multiplicity, low-symmetry $6l$ positions.

According to the generalized Neumann principle~\cite{pang2025}, the electric polarization in a crystal satisfies
\begin{equation}
(\mathcal{R}-\mathcal{I})\mathbf{p}=\mathbf{Q},
\label{eq:linalgeq}
\end{equation}
where $\mathcal{R}$ is a crystal symmetry operation, $\mathcal{I}$ is the identity, $\mathbf{p}$ is the polarization vector, and $\mathbf{Q}$ is an integer multiple of the polarization quantum. For the L1 structure, this constraint allows symmetry-allowed QFPs including $(1/3,\,2/3,\,0)$ and $(2/3,\,1/3,\,0)$. First-principles calculations give the polarization $(1/3,\,2/3,\,0)$ for L1. The L2 structure in Fig.~\ref{fig:inps3_neb}(c), related to L1 by spatial inversion, carries the opposite QFP $(2/3,\,1/3,\,0)$. The H structure in Fig.~\ref{fig:inps3_neb}(b) is the high-symmetry counterpart of L1, belonging to the $D_{3d}$ point group and space group $P\overline{3}1m$.

We construct a transition path connecting the L1 and L2 structures using the nudged elastic band (NEB) method~\cite{neb1,neb2}, which passes through the high-symmetry structure H. Along this pathway, the P atoms undergo negligible displacements, while the two In atoms remain fixed at their respective fractional coordinates. However, the site symmetry of the In atoms changes at H: in L1, they occupy the $1f$ and $1d$ Wyckoff positions, whereas in H they merge into the symmetry-degenerate $2d$ Wyckoff position. By contrast, the S atoms undergo small displacements, moving from the $6l$ Wyckoff positions in L1 to the $6k$ positions in H. The detailed Wyckoff positions of the H structure are listed in Table~S2 of the SM~\cite{SM}.

Importantly, $\ce{InPS3}$ preserves the $D_3$ symmetry throughout the transition pathway until reaching the high-symmetry structure H. According to the generalized Neumann principle, the $D_{3d}$ point group of H allows only the quantized polarization $(0,\,0,\,1/2)$, aside from the trivial zero polarization. This creates a fundamental incompatibility: as long as the system remains insulating and preserves $D_3$ symmetry, its QFP is constrained to stay at $(1/3,\,2/3,\,0)$, whereas the H structure requires a different allowed value. Therefore, a fully insulating continuous evolution is impossible. The mismatch can only be resolved through an IM transition, since QFP is no longer well defined once the system becomes metallic.

To validate this scenario, we compute the electric polarization and band gaps along the transition path. Figures~\ref{fig:inps3_neb}(d) and (e) show the evolution of the polarization and the direct and indirect band gaps, respectively. As the system evolves from L1 toward H, the in-plane quantized polarization remains fixed, in agreement with the symmetry constraint. The L1 structure has a band gap of $\sim 1.35~\mathrm{eV}$  using the HSE functional, which decreases monotonically along the path. At a critical point, the indirect band gap closes and the polarization becomes ill-defined, signaling the onset of a metallic state. At the high-symmetry H structure, the direct gap also vanishes, fully consistent with our symmetry-based analysis. Beyond H, the band gap reopens in a mirrored evolution toward L2. These results confirm the predicted interplay between symmetry and QFP along the transition pathway.

	\begin{figure}[tbp]
		\centering
		\includegraphics[width=0.45\textwidth]{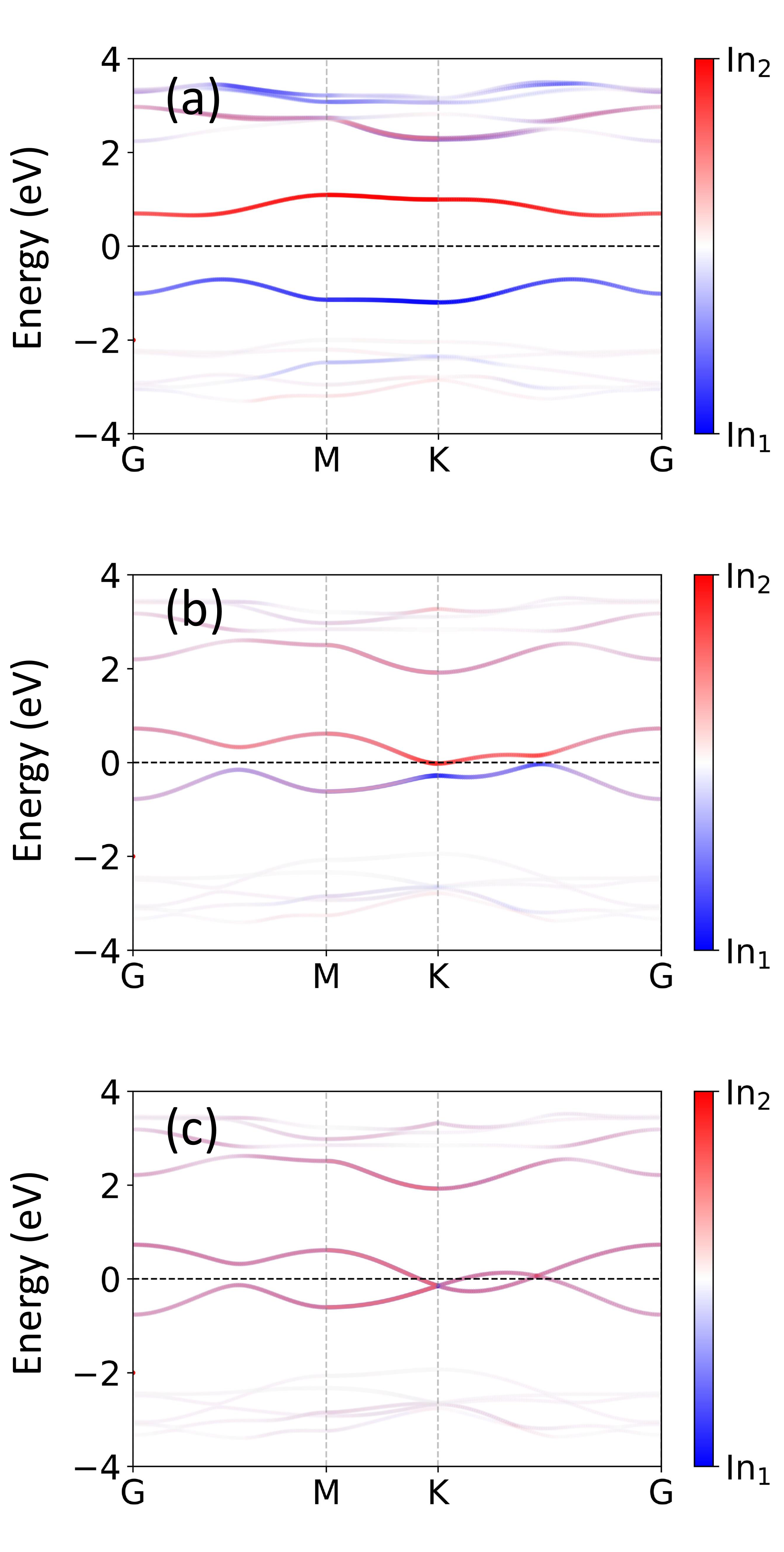}
\caption{Projected band structure of $\ce{InPS3}$ along the transition pathway. Blue and red indicate contributions from In$_1$ and In$_2$, respectively. (a) Initial state L1: the occupied states near the Fermi level are dominated by In$_1$, while the unoccupied states are mainly derived from In$_2$. (b) Intermediate state between L1 and H: the occupied and unoccupied states approach each other and exhibit mixed contributions from In$_1$ and In$_2$. (c) High-symmetry intermediate state H: the contributions from the two In atoms become identical due to symmetry-enforced degeneracy.}
		\label{fig:inps3_fatband}
	\end{figure}
	
The evolution of the electronic band structure from L1 to H is shown in Figs.~\ref{fig:inps3_fatband}(a)–(c). We focus on the two bands closest to the Fermi level, which mainly arise from the hybridization between In $s$ orbitals and the surrounding S $p$ orbitals. The color scale represents the relative contributions of the two In atoms, with blue and red denoting In$_1$ and In$_2$, respectively.

At the L1 structure, as shown in Fig.~\ref{fig:inps3_fatband}(a), the occupied states near the Fermi level are dominated by In$_1$, whereas the unoccupied states are mainly derived from In$_2$. As the system evolves along the transition path, the valence and conduction bands move toward each other, and the contributions from In$_1$ and In$_2$ become increasingly mixed, leading to a continuous reduction of the band gap. At a critical point, the indirect gap closes, as shown in Fig.~\ref{fig:inps3_fatband}(b), driving the system into a metallic state. This is accompanied by the disappearance of the electric polarization shown in Fig.~\ref{fig:inps3_neb}(d).
	
At the high-symmetry H structure, two Dirac points appear near the Fermi level, as shown in Fig.~\ref{fig:inps3_fatband}(c). One is located at K, 150 meV below the Fermi level, and is strictly protected by $D_{3d}$ symmetry. The other lies along the $\Gamma$-K line, 62 meV above the Fermi level, and is an accidental Dirac point not protected by symmetry. Both points exhibit cone-like linear dispersions. The spin-resolved Berry curvature shows a divergent feature at K, with the opposite sign at the symmetry-related $-K$ point. Further details are provided in the SM~\cite{SM}.

\begin{figure}[t]
		\centering
		\includegraphics[width=0.5\textwidth]{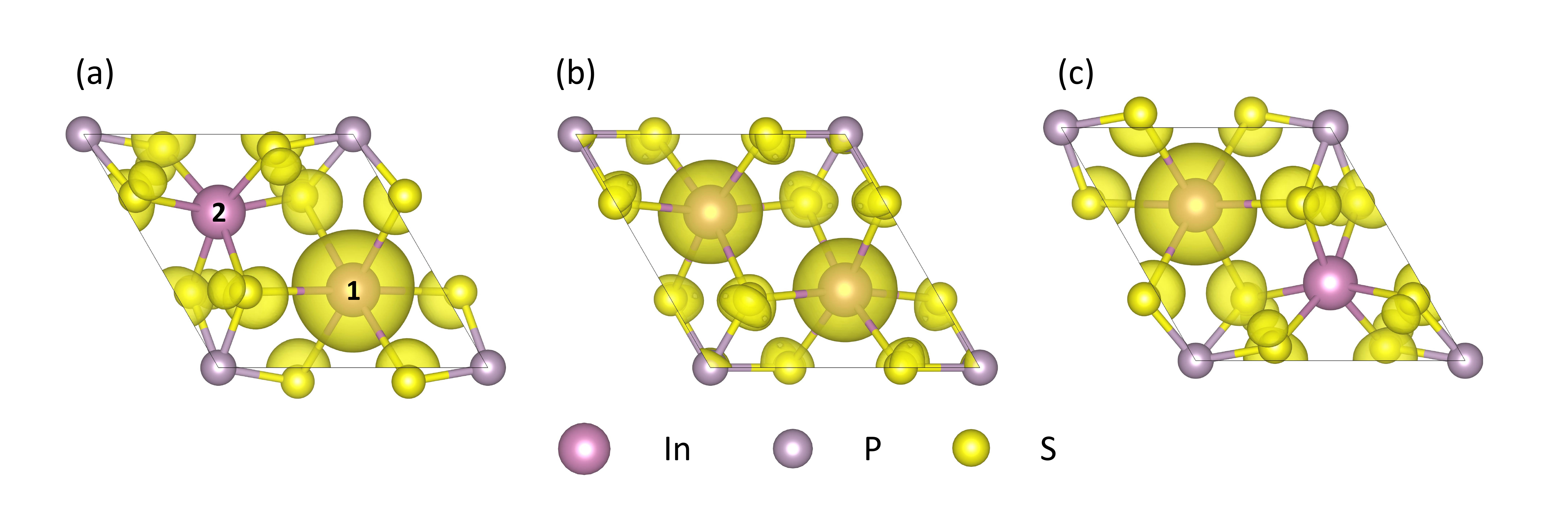}
\caption{Top view of the crystal structure and partial charge density distribution of $\ce{InPS3}$. The two In atoms are labeled as In$_1$ and In$_2$. (a) Initial state L1 with polarization $(2/3,1/3)$, where the electronic states are predominantly localized on In$_1$. (b) High-symmetry state H, where the charge density is equally distributed over the two In atoms. (c) Final state L2, where the electronic states are predominantly localized on In$_2$.}
		\label{fig:inps3_stru_pcharge}
	\end{figure}

At H, the valence and conduction bands contain strongly mixed contributions from In$_1$ and In$_2$. Beyond H, the In$_1$-derived band continues to move upward, while the In$_2$-derived band shifts downward, leading to a reopening of the gap in a manner symmetric to the evolution from L1 to H. This electronic evolution is accompanied by a continuous redistribution of charge between the two In atoms during the polarization reversal from L1 to L2, as shown in Fig.~\ref{fig:inps3_stru_pcharge}. The plotted charge density is obtained by integrating the isolated band manifold below the Fermi level.

In the initial L1 state, the electronic charge is predominantly localized on In$_1$, resulting in a pronounced charge imbalance between the two inequivalent In sites. As the system evolves along the transition path, the symmetry between the two sites is progressively restored, driving a continuous transfer of charge from In$_1$ to In$_2$. At the high-symmetry H structure, the charge is evenly distributed between the two In atoms. Beyond H, the charge continues to shift toward In$_2$ until the system reaches L2, completing the polarization reversal.
This continuous charge redistribution provides a microscopic picture of the polarization switching and its intimate connection to the insulator-to-metal transition. Additional evidence for this behavior is provided by the projected density of states (PDOS) in the SM~\cite{SM}.

Remarkably, although the atomic displacements along the L1-to-L2 pathway are extremely small, the associated change in polarization reaches as much as $2/3$ of the polarization quantum. This sharply contrasts with conventional ferroelectrics, where polarization switching is typically driven by substantial ionic displacements. Here, the dominant mechanism is instead charge transfer between the two In atoms, highlighting the essentially electronic character of the polarization reversal and suggesting a route toward large polarization responses with minimal structural distortion.

We emphasize that the NEB method\cite{neb1,neb2} serves here merely as a numerical tool to explicitly construct a continuous structural pathway between the low- and high-symmetry phases. Our conclusion does not depend on the specific transition path, but applies generally to any continuous structural evolution that preserves the symmetry of the low-symmetry phase while connecting it to a higher-symmetry structure.
Whenever such a symmetry-preserving path exists and the initial phase is insulating, QFP mismatch inevitably drives an insulator-to-metal transition.

This scenario is fundamentally different from previously studied polarization-switching mechanisms in FQFEs~\cite{Ji2024,Yu2025,pang2025quantized}, where a fully insulating path exists and the polarization remains well defined throughout the evolution. In those cases, polarization reversal proceeds through an intermediate symmetry-breaking step, for example by lowering the symmetry to $C_1$, along which the polarization is no longer quantized and can vary continuously without closing the electronic gap. Crucially, no continuous symmetry-preserving path directly connects the low- and high-symmetry phases in such examples, and the QFP-mismatch constraint identified here therefore does not apply.


	\begin{figure}[tbp]
		\centering
		\includegraphics[width=0.45\textwidth]{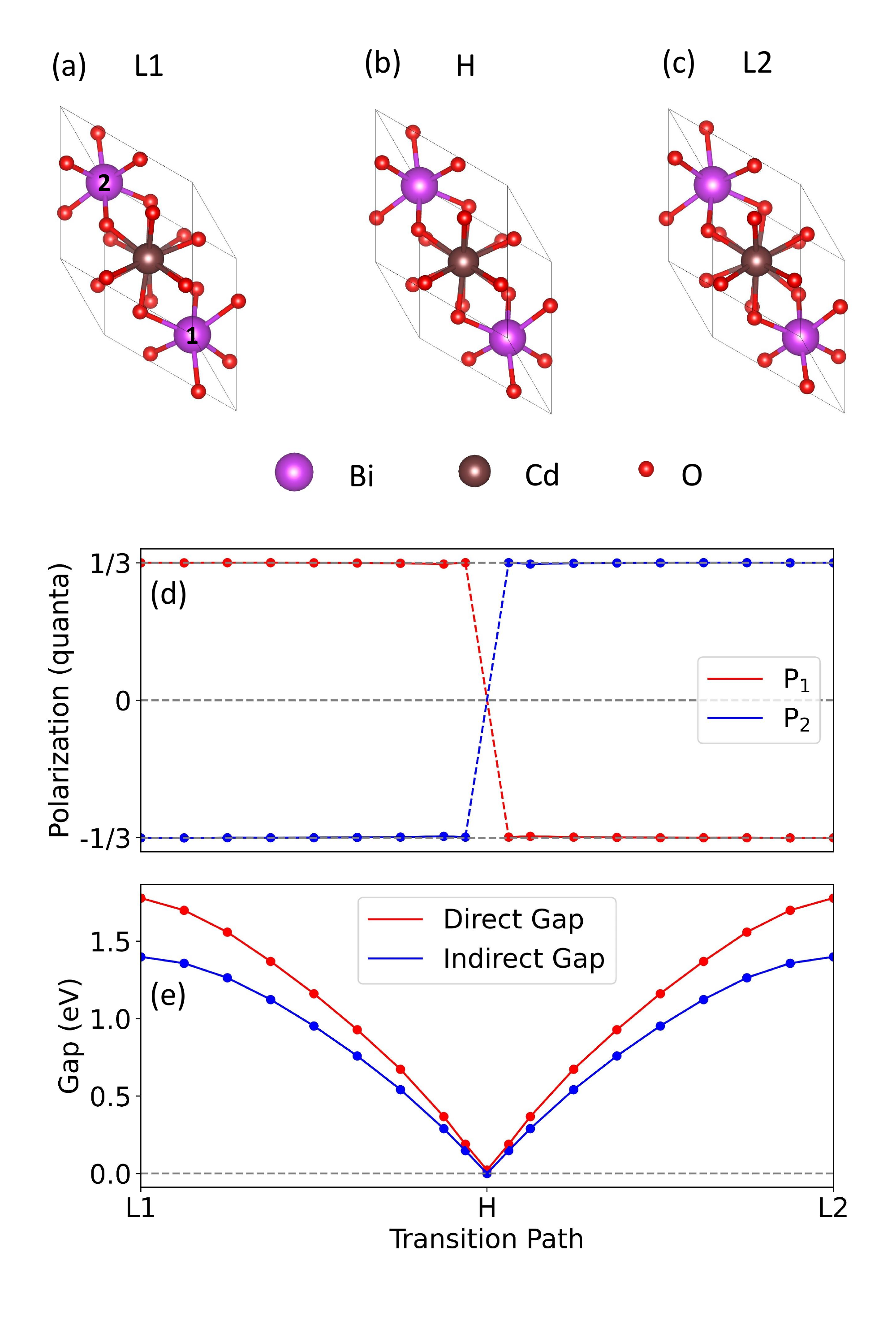}
\caption{Crystal structure of the primitive cell of $\ce{CdBiO3}$ and the evolution of its electronic properties along the transition pathway. (a)–(c) Top views of the three representative phases: (a) L1, (b) H, and (c) L2. The two Bi atoms are labeled for distinction. (d) Evolution of polarization along the transition pathway, with the dashed line indicating the metallic state. (e) Evolution of the direct and indirect band gaps along the transition pathway.}
		\label{fig:cdbio3_neb}
	\end{figure}
	
To further demonstrate the generality of our theory, we consider the three-dimensional material $\ce{CdBiO3}$ as an additional example, which exhibits a symmetry-driven IM transition enforced by the same QFP-mismatch mechanism. The low-symmetry structure, denoted as L1 in Fig.~\ref{fig:cdbio3_neb}(a), crystallizes in space group $R3$ and belongs to the point group $C_3$, while L2 in Fig.~\ref{fig:cdbio3_neb}(c) is related to L1 by spatial inversion. In L1, Cd atoms occupy the $3a$ Wyckoff positions, Bi atoms occupy another set of $3a$ positions with different internal parameters, and O atoms reside at the highly degenerate nonfractional $9z$ positions.

The corresponding high-symmetry structure, denoted as H in Fig.~\ref{fig:cdbio3_neb}(b), belongs to the $C_{3i}$ point group and crystallizes in space group $R\overline{3}$. In this structure, both Cd and Bi atoms occupy the symmetry-degenerate $6c$ positions, while O atoms move to the $18f$ positions, forming the high-symmetry intermediate configuration.
	
Applying the generalized Neumann principle to the L1 phase of $\ce{CdBiO3}$, we find that the allowed QFPs along the in-plane lattice-vector directions of the conventional cell are $(1/3,2/3)$ and $(2/3,1/3)$, while the polarization along the $c$ axis remains unrestricted. First-principles calculations confirm that the L1 phase indeed realizes the in-plane QFP $(1/3,2/3)$. The detailed symmetry analysis is presented in the SM~\cite{SM}. By contrast, the intermediate state H has $C_{3i}$ symmetry and permits only zero in-plane QFP~\cite{pang2025quantized}.

We identify a continuous path connecting L1 to H and subsequently to L2, while preserving the symmetry throughout the evolution except at the H point. Along this structural pathway, all atomic displacements remain extremely small: 0.02~\AA\ for Cd, 0.07~\AA\ for Bi, and only 0.15~\AA\ for the O atoms that dominate the modulation. These displacements are much smaller than those typically found in conventional FQFE materials.	

The evolution of polarization along the switching path is shown in Fig.~\ref{fig:cdbio3_neb}(d), and the corresponding direct and indirect band gaps are shown in Fig.~\ref{fig:cdbio3_neb}(e). Throughout the evolution, the QFP remains invariant despite the continuous reduction of both the direct and indirect band gaps. As the structure approaches H, the gap closes and the polarization disappears. These results are fully consistent with our theoretical predictions.

	\begin{figure}[tbp]
		\centering
		\includegraphics[width=0.45\textwidth]{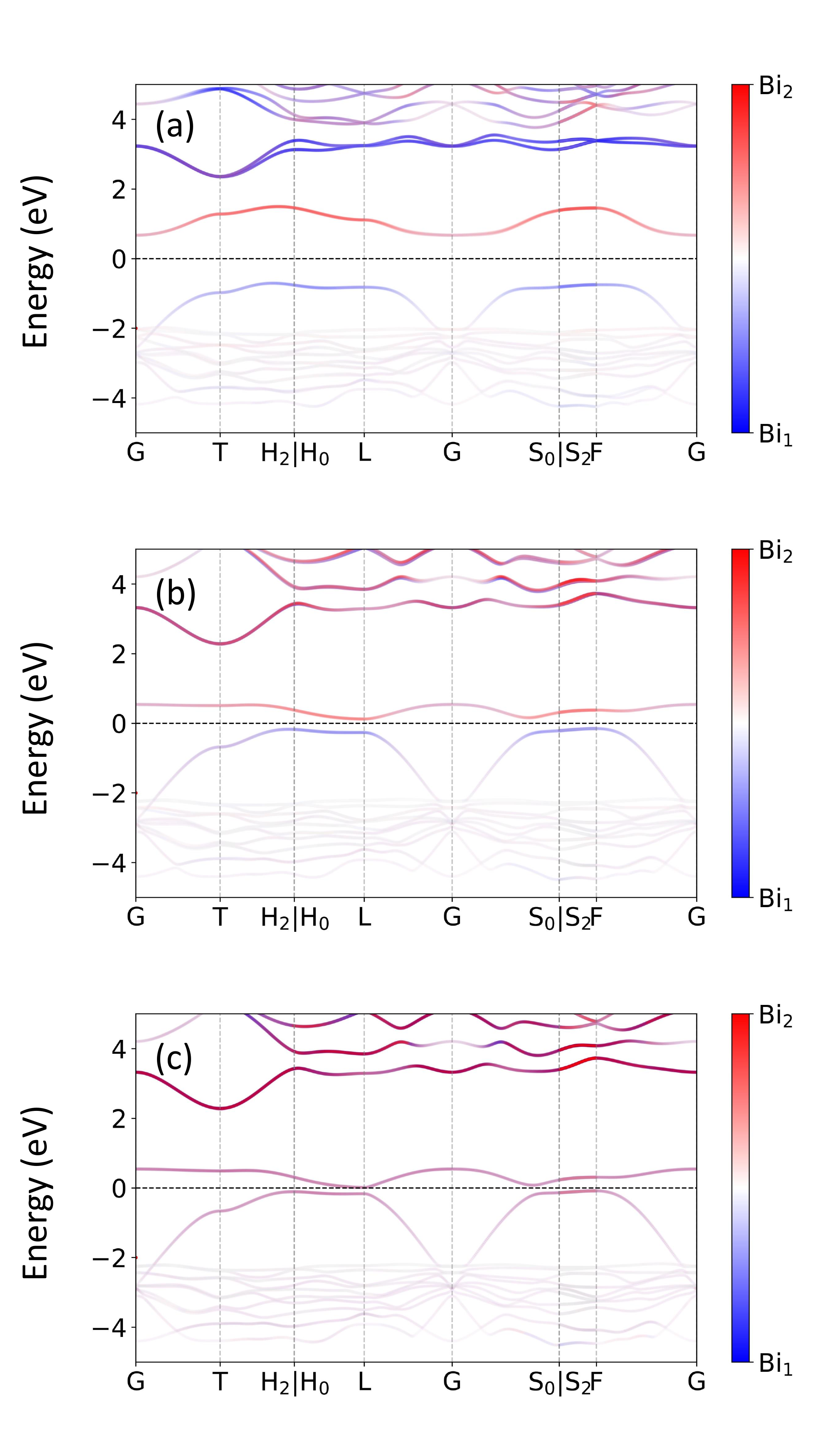}
\caption{Projected band structure of $\ce{CdBiO3}$ along the transition pathway. Blue and red indicate contributions from Bi$_1$ and Bi$_2$, respectively. (a) Initial state L1: the highest valence band is mainly derived from Bi$_1$, while the lowest conduction band is dominated by Bi$_2$. (b) Intermediate state between L1 and H: the occupied and unoccupied states approach each other, with mixed contributions from Bi$_1$ and Bi$_2$. (c) High-symmetry intermediate state H: the contributions from the two Bi atoms become identical due to symmetry-enforced degeneracy.}
		\label{fig:cdbio3_fatband}
	\end{figure}

Figure~\ref{fig:cdbio3_fatband} shows the evolution of the band structure near the Fermi level from L1 to H. The two bands closest to the Fermi level are mainly derived from the $s$ orbitals of Bi$_1$ and Bi$_2$, with the occupied and unoccupied states dominated by Bi$_1$ and Bi$_2$, respectively. As the system evolves, the hybridization between these two orbital characters strengthens as the gap decreases and becomes complete at H.

The H structure is metallic because the global indirect gap closes in the Brillouin zone, despite a finite direct gap of 15 meV. The minimum direct gap lies away from the high-symmetry lines. This picture is further supported by the finite density of states at the Fermi level. Additional details are provided in the SM~\cite{SM}.

	%
	%

In conclusion, we present a mechanism for the IM transition driven by the mismatch of QFP. Our findings offer new insights into the role of symmetry in quantum polarization and phase transitions, providing a theoretical framework for understanding similar transitions in other high-symmetry materials. This work not only deepens the understanding of symmetry-driven phase transitions but also provides a route for designing materials with tunable quantum polarization and IM transitions. In particular, the coexistence of large polarization change and strong band-gap modulation under extremely small atomic displacements makes this mechanism promising for future electronic devices with efficient switching and minimal structural distortion.

This work was supported by the Advanced Materials–National Science and Technology Major Project (Grant No. 2025ZD0618401), the National Natural Science Foundation of China (Grant No. 12134012), the Strategic Priority Research Program of the Chinese Academy of Sciences (Grant No. XDB0500201), and the Innovation Program for Quantum Science and Technology (Grant No. 2021ZD0301200). The numerical calculations were performed on the USTC High-Performance Computing facilities.	
	

%

\end{document}